\documentclass[conference]{IEEEtran}
\IEEEoverridecommandlockouts
\usepackage{cite}
\usepackage{amsmath,amssymb,amsfonts}
\usepackage{algorithmic}
\usepackage{graphicx}
\usepackage{textcomp}
\usepackage{xcolor}
\usepackage[lofdepth,lotdepth]{subfig}
\usepackage{caption}
\usepackage[inline]{enumitem}
\usepackage{amsmath}

\usepackage[nolist]{acronym} 
\usepackage{comment}
\usepackage{multirow}
\usepackage[flushleft]{threeparttable}

\def\BibTeX{{\rm B\kern-.05em{\sc i\kern-.025em b}\kern-.08em
    T\kern-.1667em\lower.7ex\hbox{E}\kern-.125emX}}
\begin{acronym}
\acro{NN}{neural network}
\acro{RNN}{recurrent neural network}
\acro{GRU}{gated recurrent unit}
\acro{LSTM}{long short-term memory}
\acro{MFCC}{Mel-frequency cepstral coefficient}
\acro{PEM}{per-epoch noise mixing}
\acro{SNR}{signal-to-noise ratio}
\acro{ASR}{automatic speech recognition}
\acro{CNN}{convolutional neural network}
\acro{DNN}{deep neural network}
\acro{FNN}{feedforward neural network}
\acro{ACCAN}{accordion annealing}
\acro{WSJ}{Wall Street Journal}
\acro{CTC}{Connectionist Temporal Classification}
\acro{WFST}{Weighted Finite State Transducer}
\acro{WER}{word error rate}
\acro{LER}{character error rate}
\acro{LVCSR}{large-vocabulary continuous speech recognition}
\acro{ROI}{range of interest}
\acro{STAN}{sensor transformation attention network}
\acro{SER}{sequence error rate}
\acro{HMM}{Hidden Markov Model}
\acro{LER}{label error rate}
\acro{sMBR}{state-level minimum Bayes risk}
\acro{MLLT}{maximum likelihood linear transform}
\acro{fMLLR}{feature-space maximum likelihood linear regression}
\acro{MVDR}{minimum variance distortionless response}
\acro{STFT}{short-time Fourier transform}
\acro{ATTACC}{attention accuracy}
\acro{ATTCORR}{attention correlation}
\end{acronym}

\usepackage{lipsum}
\usepackage{fancyhdr}

\fancypagestyle{sec}{\lhead{\tmpx}}

\fancypagestyle{arXiv}
{
   \fancyhf{}
   \chead{\textcolor{gray}{This paper has been accepted for publication at the \\
   IEEE International Conference on Artificial Intelligence Circuits and Systems (AICAS), Genoa, 2020}}
   \cfoot{\thepage}
   
}

\begin{document}

\title{EdgeDRNN: Enabling Low-latency Recurrent Neural Network Edge Inference}

\author{
    \IEEEauthorblockN{Chang Gao\IEEEauthorrefmark{1}, Antonio Rios-Navarro\IEEEauthorrefmark{2}, Xi Chen\IEEEauthorrefmark{1}, Tobi Delbruck\IEEEauthorrefmark{1}, Shih-Chii Liu\IEEEauthorrefmark{1}}
    \IEEEauthorblockA{\IEEEauthorrefmark{1}Institute of Neuroinformatics, University of Zurich and ETH Zurich, Zurich, Switzerland
    }
    \IEEEauthorblockA{\IEEEauthorrefmark{2}Robotic and Technology of Computers Lab, Universidad de Sevilla, Seville, Spain
    \\\{chang, xi, tobi, shih\}@ini.uzh.ch,
    \{arios\}@us.es}
}
\maketitle


\begin{abstract}
This paper presents a Gated Recurrent Unit (GRU) 
based recurrent neural network (RNN) accelerator called EdgeDRNN 
designed for portable edge computing. 
EdgeDRNN adopts the spiking neural network inspired delta network algorithm to 
exploit temporal sparsity in RNNs. It reduces off-chip memory access by a factor of up to 10x 
with tolerable accuracy loss.
Experimental results on a 10 million parameter 2-layer GRU-RNN, with weights stored in DRAM, show that EdgeDRNN computes them in under 0.5\,ms. 
With 2.42\,W wall plug power on an entry level USB powered FPGA board, it achieves latency comparable with a 92\,W Nvidia 1080 GPU.
It outperforms NVIDIA Jetson Nano, 
Jetson TX2 and Intel Neural Compute Stick 2 in latency by 6X. 
For a batch size of 1, EdgeDRNN achieves a mean effective 
throughput of 20.2\,GOp/s and a wall plug power efficiency that is 
over 4X higher than all other platforms.
\end{abstract}


\section{Introduction}
\thispagestyle{arXiv}
Recurrent Neural Networks (\textbf{RNN}) are a subset of deep neural networks that are particularly 
useful for regression and classification tasks involving time series inputs.
Gated RNNs which use Long Short-Term Memory units (\textbf{LSTM})~\cite{lstm_hoch97} 
and Gated-Recurrent Unit (\textbf{GRU})~\cite{gru_og} are used to  
overcome the vanishing gradient problem frequently encountered 
during RNN 
training with backpropagation through time. RNN models are frequently used in state-of-the-art models for automatic speech recognition tasks~\cite{Amodei2016,Ravanelli2019}.

In edge computing, computations are done locally on end-user devices to reduce latency and protect privacy~\cite{chen2019}. 
RNNs achieve high accuracy at the cost of large memory footprint and expensive computation. 
RNNs are usually computed on the cloud with results sent to edge devices, 
which introduces high and variable latency,
making it hard to guarantee real time performance for human computer interaction, 
robotics, and control applications. Previous work exploits weight pruning~\cite{han2017ese}~\cite{cao2019}, structured weight matrix~\cite{Wang2018}, and temporal 
sparsity~\cite{GaoDeltaRNN2018} to accelerate RNN computation 
by reducing the memory bottleneck of RNNs. 
However, these works used expensive FPGA boards with greater than 15\,W power 
consumption and did not target portable edge devices with low latency demands and a limited power budget.

This paper describes an RNN accelerator for edge applications. The accelerator exploits temporal sparsity using the delta network (\textbf{DeltaGRU})~\cite{neil2016delta} algorithm. It achieves sub-millisecond inference of big multi-layer RNNs comparable with a desktop-level GPU, 
but with 38 times less power.



\section{Gated-Recurrent Unit \& Delta Network}
\label{sec:gru}
The equations for a GRU layer of $M$ neurons and $N$-dimensional input are given as:
\begin{equation}
\begin{aligned}
\mathbf{r}_{t} &=\sigma\left(\mathbf{W}_{ir}\mathbf{x}_{t}+\mathbf{W}_{hr}\mathbf{h}_{t-1}+\mathbf{b}_{r}\right)\\
\mathbf{u}_{t} &=\sigma\left(\mathbf{W}_{iu}\mathbf{x}_{t}+\mathbf{W}_{hu}\mathbf{h}_{t-1}+\mathbf{b}_{u}\right)\\
\mathbf{c}_{t} &=\textrm{tanh}\left(\mathbf{W}_{ic}\mathbf{x}_{t}+\mathbf{r}_{t}\odot\left(\mathbf{W}_{hc}\mathbf{h}_{t-1}\right)+\mathbf{b}_{c}\right)\\
\mathbf{h}_{t} &=\left(1-\mathbf{u}_{t}\right)\odot \mathbf{c}_{t}+\mathbf{u}_{t}\odot \mathbf{h}_{t-1}
\label{eq:1}
\end{aligned}
\end{equation}

\noindent where $\mathrm{r}, \mathrm{u}, \mathrm{c}\in\mathcal{R}^{M}$ are respectively the reset gate, 
the update gate and the cell state. $W_{i}\in\mathcal{R}^{M\times N}$, $W_{h}\in\mathcal{R}^{M\times M}$
are weight matrices and $b\in\mathcal{R}^{M}$ are bias vectors. 
$\sigma$ denotes the logistic sigmoid.

Inspired by spiking neural networks, the DeltaGRU~\cite{neil2016delta} reduces operations in GRU-RNNs while maintaining high prediction accuracy. 
In DeltaGRU, 
weights are multiplied with the delta vectors $\Delta\mathbf{x}_{t}=\mathbf{x}_{t}-\mathbf{x}_{t-1}$, $\Delta\mathbf{h}_{t-1}=\mathbf{h}_{t-1}-\mathbf{h}_{t-2}$ 
between the current and the previous time steps and then added to 
a memory term $\mathbf{M}_t=\sum_{i=0}^{i=t}\left(\mathbf{W}\Delta\mathbf{x}_i+\mathbf{W}\Delta\mathbf{h}_{i-1}\right)$ 
that is the accumulation of all 
previous products. The initial states are $\mathbf{M}_0=\mathbf{b}$, 
$\mathbf{x}_{-1}=0$ and $\mathbf{h}_{-1}=\mathbf{h}_{-2}=0$. 

By setting the elements of a delta vector to zero when their individual values 
are less than a defined \textit{\textbf{Delta Threshold}} $\Theta$, the number of
 matrix-vector multiply-and-accumulate (\textbf{MAC}) operations 
 is reduced by 5X to 100X, depending on the 
 dynamics of the input and hidden units~\cite{neil2016delta}. 
 It allows skipping entire 
 columns of the weight matrix. 
 That way, DRAM weight memory reads are still in efficient burst mode.

\section{EdgeDRNN Accelerator}
\label{sec:edgedrnn}

\begin{figure}[!t]
	\centering
\subfloat[][]{  
    \includegraphics[width=.5\linewidth]{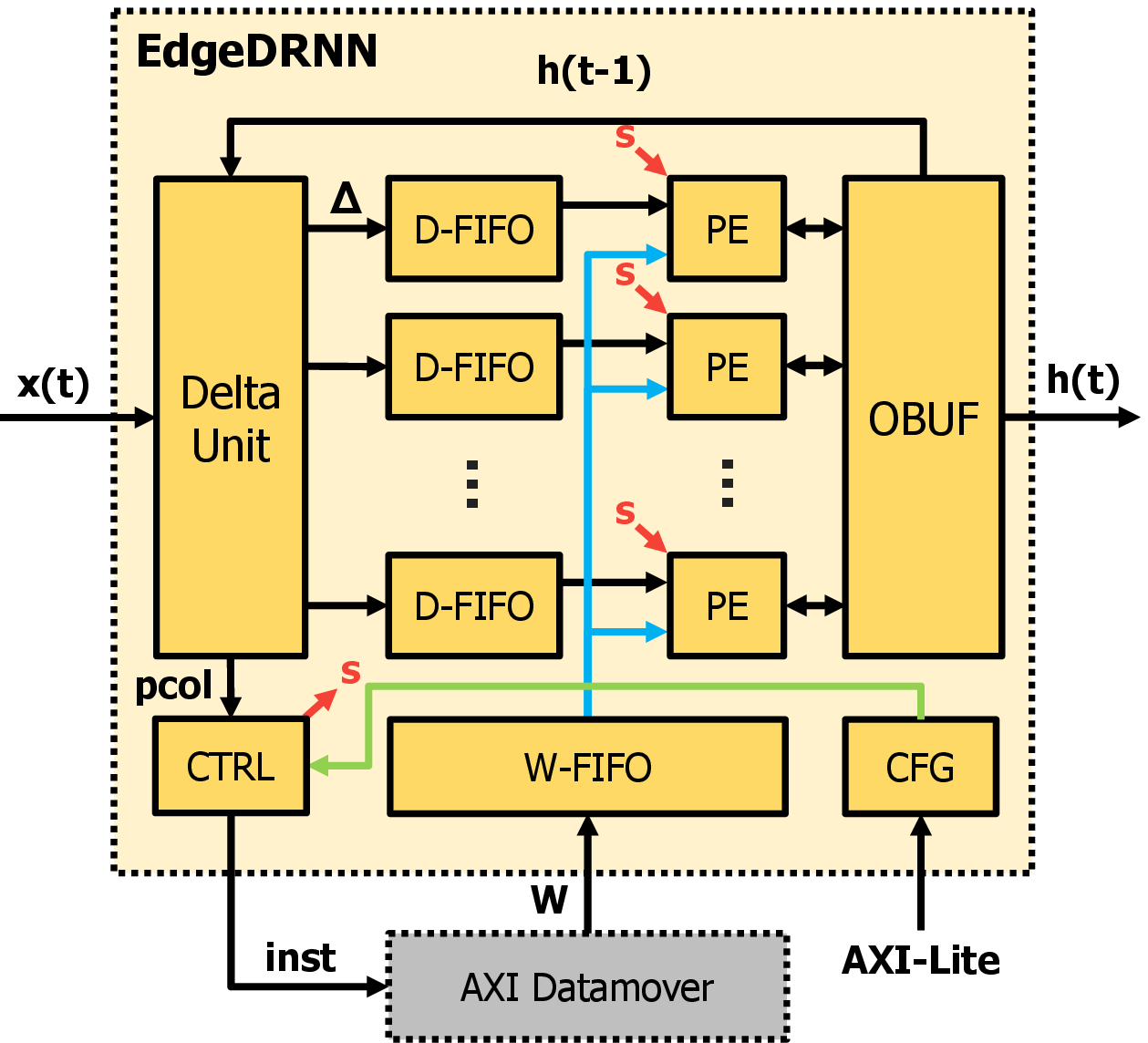}
     \label{fig:edgedrnn}
} 
\subfloat[][]{ 
    \includegraphics[width=.4\linewidth]{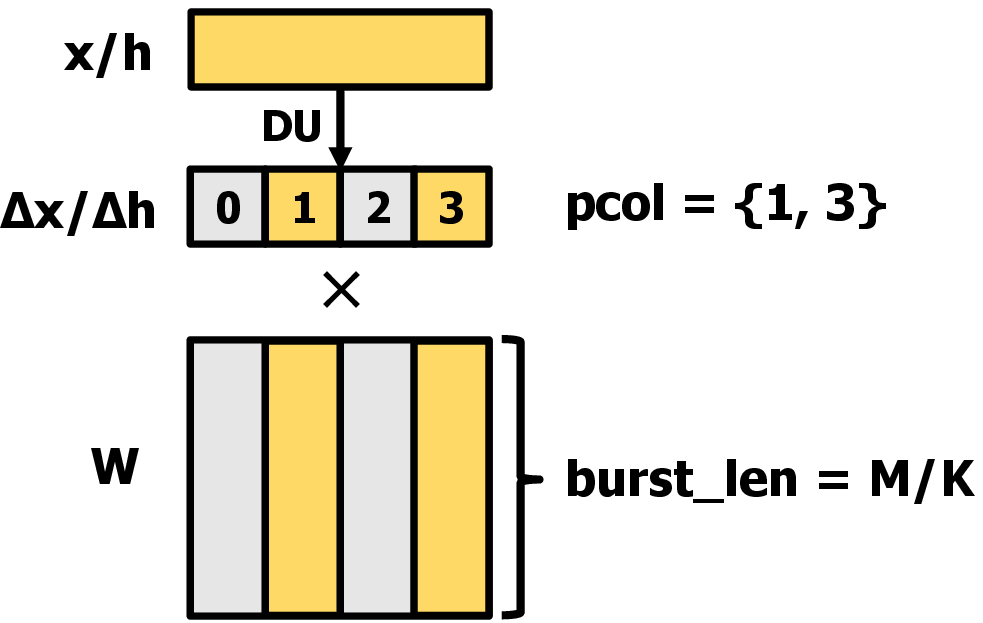}
     \label{fig:mxv}
     }
\caption{(a) EdgeDRNN accelerator architecture; (b) Flow chart of the sparse matrix-vector multiplication.}
\end{figure}

\begin{figure}[!t]
	\centering
	\includegraphics[width=0.32\textwidth]{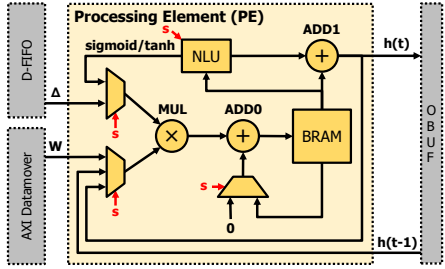}
	\caption{Architecture of the EdgeDRNN processing element (PE).}
	\label{fig:pe}
\end{figure}

\begin{figure}[!t]
	\centering
	\includegraphics[width=0.32\textwidth]{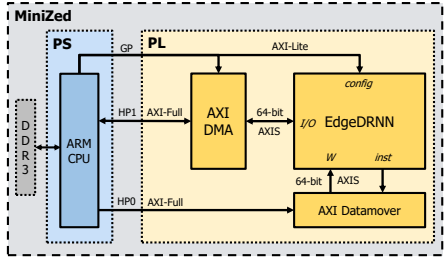}
	\caption{Top-level diagram of the EdgeDRNN implementation on the MiniZed development board.}
	\label{fig:top}
\end{figure}

\subsection{Accelerator Design}

The design of EdgeDRNN aims to achieve low-latency RNN inference with batch size of 1, 
which are needed for real-time operation with minimum latency.
2D arithmetic unit arrays are not suitable here due to 
limited weight reuse, scarce on-chip memory resources and narrow external memory
interface on embedded systems like MiniZed. 
The vector processing element (\textbf{PE}) array in EdgeDRNN is able to 
fully utilize the external memory bandwidth.

Fig.~\ref{fig:edgedrnn} shows the design of the EdgeDRNN accelerator. 
The number of PEs, $K$, in EdgeDRNN is $K=BW_{\rm DRAM}/BW_{\rm W}=64/8=8$, 
where $BW_{\rm W}=8$ is the weight precision and $BW_{\rm DRAM}=64$ the external memory interface bit-width. EdgeDRNN can be configured to support 1, 2, 4, 8, 16-bit fixed-point weights and 16-bit fixed-point activations; in this paper we used only 8-bit weights.
The delta unit (\textbf{DU}) includes BRAM memory that records previous states
$x_{t-1}$ and $h_{t-2}$ to be used for calculating delta vectors $\Delta x$ and $\Delta h$. 
The DU checks one element in a delta vector per cycle. Elements that exceed $\Theta$ result in non-zero 
elements and are broadcast
to all D-FIFOs that drive PEs. 
As shown in Figs.~\ref{fig:edgedrnn} and~\ref{fig:mxv}, 
DU computes column pointers (\texttt{pcol}) to non-zero delta vector elements 
that are sent to the global controller (\textbf{CTRL}).
Using \texttt{pcol}, 
CTRL generates instructions, containing the physical start address of
a weight column and the burst length given in Fig.~\ref{fig:mxv}, 
to control the AXI Datamover IP to fetch weights (biases are appended to weights).
On MiniZed, DRAM data moves through the PL's DMA and Datamover. 

Fig.~\ref{fig:pe} shows the design of the PE. 
The PE has a 16-bit multiplier \textbf{MUL} and two adders, 
32-bit \textbf{ADD0} and 16-bit \textbf{ADD1}. 
Multiplexers are placed before operands of MUL to reuse it 
in both matrix-vector multiplications between delta vectors $\Delta$ and weights $W$, 
and any element-wise multiplication. 
The nonlinear unit (\textbf{NLU}) uses look-up tables (\textbf{LUT}) to compute quantized \texttt{sigmoid} and \texttt{tanh} functions.
The multiplexer below ADD0 selects between BRAM data and '0' for accumulation 
and necessary BRAM initialization respectively. 
Signal $s$ from CTRL is used to control multiplexers and select target nonlinear function of NLU.
ADD1 is responsible for element-wise additions and 
sends the output activation $h$ to output buffer \textbf{OBUF}.

\begin{table}[!t]
\caption{Resource utilization of MiniZed.}
\label{tab:resource}
\centering
\resizebox{0.46\textwidth}{!}{ 
\begin{tabular}{|l|c|c|c|c|c|}
\hline
\textbf{}           & \textbf{LUT} & \textbf{LUTRAM} & \textbf{FF} & \textbf{BRAM} (36Kb) & \textbf{DSP} \\ \hline
\textbf{Available}  & 14400        & 6000            & 28800       & 50            & 66           \\ \hline
\textbf{Used}       & 10464        & 552             & 11665       & 33            & 9            \\ \hline
\textbf{Percentage} & 72.67\%      & 9.20\%          & 40.50\%     & 66\%          & 13.64\%      \\ \hline
\end{tabular}
}
\end{table}

\begin{figure}[!t]
	\centering
	\includegraphics[width=0.48\textwidth]{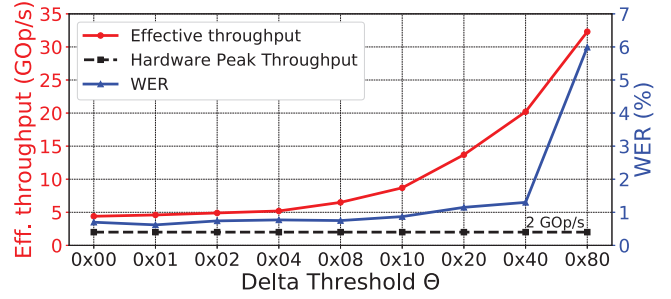}
	\caption{Mean effective throughput and word error rate evaluated on the \texttt{TIDIGITS} test set versus various delta thresholds (shown as hex values corresponding to 0$\sim$0.5 floating point threshold) used in both training and inference of a 2L-768H-DeltaGRU network.}
	\label{fig:delta}
\end{figure}
\begin{table}[!t]
\caption{Word error rate (WER) of GRU and DeltaGRU networks trained with $\Theta=0$x$40$, $\beta=$1e-5 on TIDIGITS.}
\label{tab:wer}
\centering
\begin{tabular}{|c|c|c|c|c|}
\hline
Network Size & \#Param. & \begin{tabular}[c]{@{}c@{}}WER\\ (GRU)\end{tabular} & \begin{tabular}[c]{@{}c@{}}WER\\ (DeltaGRU)\end{tabular} & Degradation \\ \hline
1L-256H      & 0.23 M   & 1.83\%                                              & 3.19\%                                                   & +1.36\%     \\ \hline
2L-256H      & 0.62 M   & 1.13\%                                              & 1.83\%                                                   & +0.69\%     \\ \hline
1L-512H      & 0.85 M   & 1.04\%                                              & 1.49\%                                                   & +0.44\%     \\ \hline
2L-512H      & 1.86 M   & 0.89\%                                              & 1.64\%                                                   & +0.75\%     \\ \hline
1L-768H      & 2.42 M   & 1.27\%                                              & 1.38\%                                                   & +0.11\%      \\ \hline
2L-768H      & 5.40 M   & 0.77\%                                              & 1.30\%                                                   & +0.53\%     \\ \hline
\end{tabular}
\end{table}

\begin{table*}[]
\caption{Latency and throughput of EdgeDRNN on DeltaGRU networks trained with $\Theta=0$x$40$, $\beta=$1e-5.}
\label{tab:benchmark}
\centering
\begin{tabular}{|c|c|c|c|c|c|c|c|c|}
\hline
\multirow{2}{*}{Network Sizes} & \multirow{2}{*}{\begin{tabular}[c]{@{}c@{}}Op\\ (Timestep)\end{tabular}} & \multicolumn{2}{c|}{Latency ($\mu$s)} & \multicolumn{2}{c|}{Effective Throughput (GOp/s)} & \multirow{2}{*}{\begin{tabular}[c]{@{}c@{}}MAC \\ Efficiency\end{tabular}} & \multirow{2}{*}{\begin{tabular}[c]{@{}c@{}}Sparsity \\ $\Gamma_{\Delta x}$\end{tabular}} & \multirow{2}{*}{\begin{tabular}[c]{@{}c@{}}Sparsity\\ $\Gamma_{\Delta h}$\end{tabular}} \\ \cline{3-6}
                               &                                                                          & Mean (min, max)           & Est.      & Mean (min, max)                & Est.             &                                                                            &                                                                                          &                                                                                         \\ \hline
1L-256H                        & 0.45 M                                                                   & 46.4 (16.5, 142.4)        & 43.3      & 9.8 (3.2, 27.5)                & 10.5             & 490\%                                                                      & 25.6\%                                                                                   & 90.0\%                                                                                  \\ \hline
2L-256H                        & 1.24 M                                                                   & 91.0 (29.3, 259.1)        & 91.6      & 13.6 (4.8, 42.4)               & 13.6             & 682\%                                                                      & 78.9\%                                                                                   & 89.1\%                                                                                  \\ \hline
1L-512H                        & 1.70 M                                                                   & 130.7 (40.8, 331.2)       & 129.8     & 13.0 (5.1, 41.6)               & 13.1             & 649\%                                                                      & 25.6\%                                                                                   & 89.5\%                                                                                  \\ \hline
2L-512H                        & 3.72 M                                                                   & 252.8 (57.2, 657.0)       & 262.9     & 19.2 (7.4, 84.6)               & 18.4             & 958\%                                                                      & 85.5\%                                                                                   & 91.2\%                                                                                  \\ \hline
1L-768H                        & 4.84 M                                                                   & 224.3 (64.3, 616.8)       & 224.8     & 16.6 (6.0, 57.9)               & 16.6             & 830\%                                                                      & 25.6\%                                                                                   & 91.3\%                                                                                  \\ \hline
2L-768H                        & 10.80 M                                                                  & 535.7 (96.6, 1344.7)      & 541.6     & 20.2 (8.0, 111.8)              & 19.9             & 1008\%                                                                     & 87.0\%                                                                                   & 91.6\%                                                                                  \\ \hline
\end{tabular}
\end{table*}

\begin{figure}[!t]
	\centering
	\includegraphics[width=0.48\textwidth]{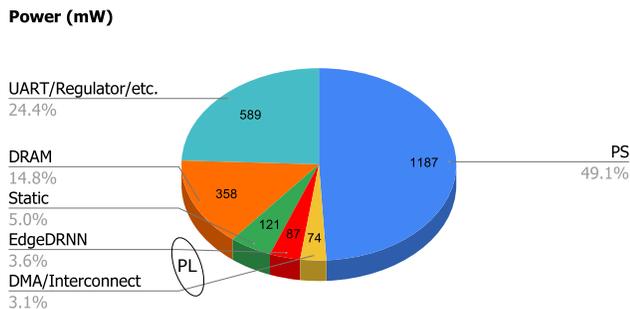}
	\caption{EdgeDRNN power breakdown on MiniZed.}
	\label{fig:power}
\end{figure}

\subsection{Implementation on MiniZed}
Fig.~\ref{fig:top} shows the implementation of EdgeDRNN on the 
Zynq-7007S system-on-chip (SoC) on the \$89  
MiniZed development board~\cite{minized}.
EdgeDRNN is implemented in the programmable logic (\textbf{PL}). 
I/O is managed by an AXI Direct Memory Access (\textbf{DMA}) IP. 
The AXI Datamover fetches weights from DDR3 memory on the Processing System (\textbf{PS}) 
side through an 64-bit ($BW_{DRAM}$) AXI-Full High Performance (\textbf{HP}) slave port. 
The AXI-Lite General Purpose (\textbf{GP}) master port is used for the single-core 
ARM Cortex-A9 CPU to control the DMA and write configurations,
including network size, delta threshold and offset address of weights, to the EdgeDRNN.
The PL is globally driven by a 125\,MHz clock from the PS. 

Table~\ref{tab:resource} shows the resource utilization of the PL. 
BRAMs are used to synthesize previous state memory in DU, 
accumulation memory in PE and FIFOs. 8 DSPs are used for the MAC units 
in 8 PEs while the remaining DSP in CTRL produces weight column addresses.
The most consumed resources are LUTs (72\%). 

\section{Experimental Results}
\label{sec:results}

We trained 6 different sizes of GRU and corresponding DeltaGRU networks 
 to compare their word error rate (\textbf{WER}) on the \texttt{TIDIGITS} audio digit dataset, evaluated using the greedy decoder. 
Inputs of all networks are 40-dimensional log filter bank features extracted 
from audio sampled at 20\,kHz and framed with 25\,ms frame size and 10\,ms frame stride. 
Networks are trained for 50 epochs using the Connectionist Temporal Classification 
(\textbf{CTC}) loss function~\cite{Graves2006} and L1 regularizer with factor $\beta$=1e-5~\cite{neil2016delta}. 
The \texttt{Adam} optimizer was used to update network parameters 
with learning rate of 3e-4 and batch size of 32. 
EdgeDRNN was configured to use INT16 activations and INT8 weights and these networks were trained in PyTorch 1.2.0 with a quantization method similar to~\cite{mishra2018wrpn}.
We used DeltaGRU  $\Theta$ from $0$ to $0.5$ (0x80). 
Training was coded in Python with PyTorch 1.2.0 and 
ran on an NVIDIA GTX 1080 GPU with CUDA\,10 and cuDNN\,7.6. 
Latency and throughput of EdgeDRNN were evaluated on DeltaGRU networks of different sizes using the first 10,000 timesteps of the test set. 
The latency is the elapsed time from when input data 
is fetched for RNN computation to when RNN output data is available in DRAM.

\begin{figure*}[!t]
	\centering
	\includegraphics[width=0.95\textwidth]{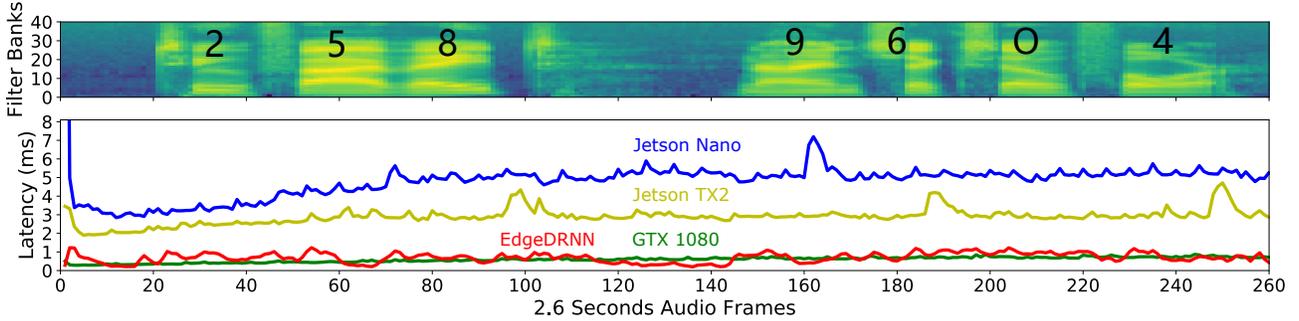}
	\caption{(Top) Audio spectrogram filter bank features with annotated labels and (bottom) measured hardware latency per frame  of a sample (25896O4A.WAV) from the \texttt{TIDIGITS} test set benchmarked on different hardware platforms.}
	\label{fig:latency}
\end{figure*}

\begin{table*}[]
\caption{Comparison of EdgeDRNN with previous work and commercial products (the 5\,W Google Edge TPU does not support RNNs).}
\label{tab:compare}
\centering
\resizebox{1.00\textwidth}{!}
{
\begin{tabular}{|l|c|c|c|c|c|c|c|c|c|}
\hline
\multirow{2}{*}{\textbf{Platform}}                                                             & \multicolumn{5}{c|}{\textbf{FPGA}}                                                                    & \textbf{ASIC}           & \multicolumn{3}{c|}{\textbf{GPU}}                                           \\ \cline{2-10} 
                                                                                               & \multicolumn{3}{c|}{\textbf{This Work}}       & \textbf{DeepStore} & \textbf{ESE}                     & \textbf{NCS2}           & \textbf{Jetson Nano}    & \textbf{Jetson TX2}     & \textbf{GTX 1080}       \\ \hline
\textbf{Chip}                                                                                  & \multicolumn{3}{c|}{XC7Z007S}                 & XC7Z045            & XCKU060                          & Myriad X                & Tegra X1                & Tegra X2                & GP104                   \\ \hline
\textbf{Dev. Kit Cost}                                                                         & \multicolumn{3}{c|}{\$89}                     & \$2,495            & \$3,295                          & \$69                    & \$99                    & \$411                   & \$500+PC                \\ \hline
\textbf{Bit Precision (A/W)}                                                                   & \multicolumn{3}{c|}{INT 16/8}                 & INT 16/16          & INT 16/12                        & FP 16/16                & FP 32/32                & FP 32/32                & FP 32/32                \\ \hline
\textbf{Test Network}                                                                          & \multicolumn{3}{c|}{DeltaGRU}                 & LSTM               & \multicolumn{1}{l|}{Google LSTM} & LSTM                    & GRU                     & GRU                     & GRU                     \\ \hline
\textbf{Network Size}                                                                          & \multicolumn{3}{c|}{2L-768H}                  & 2L-128H            & 1L-1024H                         & 2L-664H                 & 2L-768H                 & 2L-768H                 & 2L-768H                 \\ \hline
\textbf{\#Parameters}                                                                          & \multicolumn{3}{c|}{5.40 M}                   & 0.26 M             & 3.25 M                           & 5.40 M                  & 5.40 M                  & 5.40 M                  & 5.40 M                  \\ \hline
\multirow{2}{*}{\textbf{WER on TIDIGITS}}                                                      & $\Theta=0$x$00$ & $\Theta=0$x$08$ & $\Theta=0$x$40$ & \multirow{2}{*}{-} & \multirow{2}{*}{-}               & \multirow{2}{*}{1.07\%} & \multirow{2}{*}{0.77\%} & \multirow{2}{*}{0.77\%} & \multirow{2}{*}{0.77\%} \\ \cline{2-4}
                                                                                               & 0.69\%        & 0.75\%        & 1.30\%        &                    &                                  &                         &                         &                         &                         \\ \hline
\textbf{Latency ($\mu$s)}                                                                      & 2633          & 1673          & 536           & -                  & -                                & 3,588                   & 5,327                   & 3,240                   & 715                     \\ \hline
\textbf{\begin{tabular}[c]{@{}l@{}}Batch-1\\ Throughput (GOp/s)\end{tabular}}                  & 4.10          & 6.46          & 20.16         & 1.04               & 79.20                            & 3.01                    & 2.03                    & 3.33                    & 15.10                   \\ \hline
\textbf{On-Chip Power (W)}                                                                     & \multicolumn{3}{c|}{1.48}                     & 2.30               & -                                & -                       & -                       & -                       & -                       \\ \hline
\textbf{\begin{tabular}[c]{@{}l@{}}Batch-1 On-Chip \\ Power Efficiency (GOp/s/W)\end{tabular}} & 3.20          & 4.36          & 13.62         & 0.45               & -                                & -                       & -                       & -                       & -                       \\ \hline
\textbf{Wall Plug Power (W)}                                                                   & \multicolumn{3}{c|}{2.42}          & -                  & 41.00+PC                         & 1.74                    & 7.56                    & 11.70                   & 92.43+PC                \\ \hline
\textbf{\begin{tabular}[c]{@{}l@{}}Batch-1 System \\ Power Efficiency (GOp/s/W)\end{tabular}}  & 1.70          & 2.68          & 8.35          & -                  & 1.93                             & 1.73                    & 0.27                    & 0.28                    & 0.16                    \\ \hline
\end{tabular}
}
\end{table*}

\subsection{Accuracy and Throughput}
Figure~\ref{fig:delta} shows the EdgeDRNN throughput and WER versus the $\Theta$ 
used in training and testing of a 2L-768H-DeltaGRU network.
With 8\,PEs at 125\,MHz, EdgeDRNN has a theoretical peak throughput of 2\,GOp/s.
At $\Theta=0$, there is still a speedup of about 2X from natural sparsity of the delta vectors.
Higher $\Theta$ leads to better effective throughput, but with gradual slight WER degradation.
The optimal point is at $\Theta=0$x$40$ (0.25), just before a dramatic increase of WER,
where EdgeDRNN achieves an effective throughput around 20.2\,GOp/s with 1.3\% WER. 
We use the same $\Theta=0$x$40$ to train all other DeltaGRU networks 
and their accuracy is compared with GRU networks of the same size in Table~\ref{tab:wer}.
The smallest network 1L-256H-DeltaGRU has a 1.36\% WER increase.
The largest network 2L-768H-DeltaGRU achieves a 0.53\% higher WER but 4X more effective throughput.
Setting $\Theta<=0$x$08$ shows that INT16/INT8 arithmetic achieves the same accuracy 
as FP32 (Table~\ref{tab:compare}), but here the effective throughput is reduced to 6.5 versus 20.2\,GOp/s/W. 


\subsection{Theoretical \& Measured Performance}
The theoretical estimated mean effective throughput $\nu$ of EdgeDRNN running a DeltaGRU layer is given as:
\begin{align}
\nu&=\frac{\mathrm{Op}}{\tau_{M}+\tau_{A}}
\\&\approx\frac{2\left(3MN+3M^2(L-1)+3M^2L\right)}{\frac{\left(3MN+3M^2(L-1)\right)(1-\Gamma_{\Delta x})+3M^2L(1-\Gamma_{\Delta h})}{Kf}+\frac{3M}{Kf}}
\label{eq:5}
\end{align}
where $\mathrm{Op}$ is the number of operations 
in a DeltaGRU layer per timestep, $\tau_M$ the 
latency of MxV and $\tau_A$ the latency of remaining operations to produce the activation. 
$\Gamma_{\Delta x}$ and $\Gamma_{\Delta h}$ are the mean sparsity of input and hidden units respectively, $L$ the number of hidden layers and $f$ the clock frequency.

Table~\ref{tab:benchmark} compares benchmark results of different sizes of DeltaGRU networks on EdgeDRNN. 
Estimated results by Eq.~\ref{eq:5} are within 7.1\% relative error to measured results,  
so Eq.~\ref{eq:5} is useful to estimate EdgeDRNN performance. 
On average, EdgeDRNN can run all tested networks with less 
than 0.54\,ms latency, which corresponds to 20.2 GOp/s effective throughput for the 2L-768H-DeltaGRU.

\subsection{Power Measurement}

Fig.~\ref{fig:power} shows the power breakdown of the MiniZed system. The total power is measured by a USB power meter; 
the PS, PL and static power is estimated by the Xilinx Power Analyzer. 
The whole system active burns at most 2.416\,W. 
The EdgeDRNN logic burns only 87\,mW. 
Thus the wall plug and incremental power efficiency 
are 8.4\,GOp/s/W and 231.7\,GOp/s/W respectively. 
Varying modes of operation allows inferring 
EdgeDRNN DRAM memory power of 358\,mW, 
resulting in EdgeDRNN+DRAM power efficiency of 
38.3\,GOp/s/W. 
We used the wall plug power efficiency for the following comparisons.

\section{Conclusion}
\label{sec:conclude}

Table~\ref{tab:compare} compares EdgeDRNN with other platforms. 
The same task (first 10,000 timesteps of the test set) was benchmarked on EdgeDRNN, ASIC and GPUs. 
The Intel Compute Stick 2 (NCS2) 
does not support GRU and was benchmarked with an LSTM 
network with similar parameter count and trained on the same 
dataset and hyperparameters. 
For benchmark of GPUs, we used the cuDNN implementation of GRU
that achieved 715\,$\mu$s latency on NVIDIA GTX 1080, 
which is 2.4X quicker than the DeltaGRU using the NVIDIA cuSPARSE library.
We also compare this work to reported specifications
of DeepStore~\cite{deepstore}, which  has similar power 
consumption as EdgeDRNN, and ESE~\cite{han2017ese}, which is a sparse matrix-vector multiplication accelerator for LSTM.

The power efficiency results show that 
EdgeDRNN achieves over 4.8X higher system power 
efficiency compared to commercial ASIC and GPU products, 
30X higher on-chip power efficiency compared to~\cite{deepstore} 
and 4.3X higher system power efficiency than ESE.

Fig.~\ref{fig:latency} compares the latencies on a test set sample. 
EdgeDRNN is as quick as 1080 GPU and 6X quicker than the other platforms.
EdgeDRNN latency is lower during the silent or quieter periods (e.g. between 120\,s and 140\,s).



The delta threshold $\Theta$ allows instantaneous tradeoff of accuracy versus latency.
Using sparsity in delta vectors allows the arithmetic units 
on this task to effectively compute ten times more operations.

The throughput of commercial edge devices 
on batch-1 RNNs 
are a factor of more than 100X less than the claimed 
peak performance offered by these platforms, 
which range from 500\,GOp/s for Jetson Nano~\cite{jetson_nano}
up to nearly 10\,TOp/s for GTX\,1080~\cite{gtx1080}
It shows that an optimized RNN platform 
can do better in throughput and especially power efficiency. 


 \section{Acknowledgement}
 This work was partially funded by the Samsung Advanced Institute of Technology and the Swiss National Science Foundation, HEAR-EAR, 200021\_172553 grant. 



\bibliographystyle{IEEEtran}
\bibliography{IEEEabrv}

\end{document}